\documentclass[11pt]{article}
\def\withcolors{1}
\def\withnotes{0}

\usepackage{ccanonne}
\usepackage{palatino}
\newcommand{\membits}{\textcolor{orange}m}
\newcommand{\nss}{\textcolor{red}s}
\newcommand{\freq}{\textcolor{red}N}
\newcommand{\nst}{\textcolor{ForestGreen}T}

\title{Simpler Distribution Testing with Little Memory}
\author{Cl\'ement L. Canonne\thanks{University of Sydney. Email: \email{clement.canonne@sydney.edu.au}. Supported by an ARC DECRA (DE230101329) and an unrestricted gift from Google Research.} \and Joy Qiping Yang\thanks{University of Sydney. Email: \email{qyan6238@uni.sydney.edu.au}. Supported by a JD.com Fellowship.}}
  
\begin{document}
\maketitle
\begin{abstract}
    We consider the question of distribution testing (specifically, uniformity and closeness testing) in the streaming setting, \ie under stringent memory constraints. We improve on the results of Diakonikolas, Gouleakis, Kane, and Rao (2019) by providing considerably simpler algorithms, which remove some restrictions on the range of parameters and match their lower bounds.
\end{abstract}

\section{Introduction}

Distribution testing, a subfield of property testing dating to~\cite{GGR:98}, and whose systematic study was initiated in~\cite{BFRSW:00}, is concerned with making fast decisions about the statistical properties of datasets, given very few samples. As such, it is deeply related to the field of (composite) hypothesis testing in Statistics and information theory, but with a specific focus on the finite-sample regime, and in particular from a sample complexity viewpoint: what is the minimum number of observations needed to efficiently decide, with high probability of success, whether the data distribution exhibits some particular property of interest?

In this paper, we will focus on the simplest and most fundamental distribution testing task, that of \emph{uniformity testing}: given $\ns$ samples from an unknown probability distribution over a known discrete domain of size $\ab \gg 1$, and a distance parameter $\dst \in(0,1]$, decide with high constant probability whether $\p$ is \emph{the} uniform distribution $\uniform_{\ab}$ on the domain, or if it is statistically far (\ie at total variation distance at least $\dst$) from $\uniform_{\ab}$. This question has been, of course, extensively studied over the past two decades, and is by now well understood: in particular, we refer the readers to~\cite{Canonne:15:Survey,BW:17} and~\cite[Chapter~11]{Goldreich:17} for surveys on distribution testing, and to~\cite{CanonneTopicsDT2022} for a recent monograph more specifically on uniformity testing and related problems.

In spite of this extensive work on uniformity testing, not everything is so clear or settled when it comes to testing \emph{under additional constraints}, for instance when the data is sensitive (\ie testing under various privacy constraints~\cite{CDK:17,ADR:17,ASZ:18:DP,AcharyaCT:IT1,AJM:20,BB:20,AcharyaCFST21,BCJM:20,CanonneL22}), distributed (communication constraints~\cite{AndoniMalkinNosatzki:18,FMO:18,ACT:19,ACLST:20}), or~--~as will be the focus of this work~--~observed in a streaming fashion by a memory-limited device~\cite{DiakonikolasGKR19,BergOS22}.

\paragraph{Setting.} In the (one-pass) streaming setting, $\ns$ \iid samples from an unknown probability distribution $\p$ over $[\ab] \eqdef\{1,2,\dots,\ab\}$ are sequentially observed, in random order, by a memory-limited algorithm which can only keep in memory $\membits$ bits at any given time (and may, or may not, be randomized). The algorithm is provided with the parameters of the problem, namely the domain size $\ab$ and distance parameter $\dst$ (as well as the values of $\ns,\membits$), and must, at the end of the stream, output either $\accept$ or $\reject$:
\begin{itemize}
    \item if $\p=\uniform_{\ab}$, the algorithm must output $\accept$ with probability at least $2/3$;
    \item if $\totalvardist{\p}{\uniform_{\ab}}>\dst$, the algorithm must output $\reject$ with probability at least $2/3$;
\end{itemize}
where $\totalvardist{\p}{\q} = \sup_{S\subseteq [\ab]} \Paren{\p(S)-\q(S)} = \frac{1}{2}\sum_{i=1}^{\ab} \abs{\p(i)-\q(i)}$ denotes the \emph{total variation distance} (a.k.a.\ statistical distance) between two distributions $\p,\q$ over the same domain. Note that it is a promise problem: if $\p$ satisfies neither of the two conditions, then the algorithm is off the hook and can output whatever value it pleases.

We will also consider, at some point, the \emph{closeness testing} problem, a generalization where the algorithm is provided with \emph{two} streams of \iid samples, coming from two unknown distributions $\p,\q$ and must similarly distinguish at the end between $\p=\q$ and $\totalvardist{\p}{\q} > \dst$.

\paragraph{Regime of parameters.} From the above, we require that the available memory $\membits$ must be enough to at least store the value of $\ab$, and that of $\dst$ and $\ns$. This makes sense, as even accessing the current element in the stream requires reading $\log_2\ab$ bits of memory; and the algorithm should be able to keep track of how many samples have been received so far, which takes $\log_2\ns$ bits. Moreover, a lower bound of~\cite{BergOS22} shows that, even with no restriction on $\ns$ at all, at least $\Omega(\max(\log_2\ab,\log(1/\dst)))$ bits of memory are necessary for any uniformity testing algorithm.

On the other hand, we will restrict ourselves to the setting where (1)~$\membits \leq \ab\log \ns$, otherwise, one can just keep the ``counts'' (frequency of each domain element) in memory, which is a sufficient statistic; and (2)~$\membits \leq \ns\log \ab$, as otherwise the algorithm can just store all samples in memory. Thus, we will in this paper focus on the remaining ``interesting'' regime, 
\begin{equation}
    \label{eq:relevant:regime:m}
    \max(\log\ns, \log\ab, \log(1/\dst)) \leq \membits \leq \min(\ab\log\ns, \ns\log\ab)
\end{equation}

\paragraph{Prior work.} Absent memory constraints, the optimal sample complexity of uniformity testing is known to be $\bigTheta{\sqrt{\ab}/\dst^2}$~\cite{Paninski:08}, while that of closeness testing is $\bigTheta{\max\Paren{\sqrt{\ab}/\dst^2,\ab^{2/3}/\dst^{4/3}}}$~\cite{ChanDVV14}. In the streaming setting, the study of uniformity testing was initiated by Diakonikolas, Gouleakis, Kane, and Rao~\cite{DiakonikolasGKR19}, who prove both upper and lower bounds on the trade-off between $\ns$ and $\membits$ for uniformity testing, as well as an upper bound for closeness testing. Their results build on a specific uniformity testing algorithm they propose and analyze, the \emph{bipartite collision tester}, which they then leverage for distribution testing in both the streaming setting and a (specific) communication-limited setting. However, the analysis of their bipartite collision tester is quite involved (spanning roughly five pages), and comes with some inherent limitations on the range of parameters allowed. We summarize their results in~\cref{table:sample_complexity_uniformity,table:sample_complexity_closeness}.

\begin{table}[ht]
  \centering
  \begin{tabular}{|c|c|c|c|c|}
    \hline
    & \multicolumn{4}{c|}{\textbf{Existing Sample Complexity Bounds }}\\
    \hline
    \multirow{2}*{\textbf{Property}} & {\textbf{Upper Bound}} & {\textbf{Lower Bound
    1}} & {\textbf{Lower Bound 2}} & {\textbf{Lower Bound 3}} \\
     & {\cite{DiakonikolasGKR19}} & {\cite{DiakonikolasGKR19}} &
    {\cite{DiakonikolasGKR19}} & {\cite{BergOS22}}\\
    \hline
    Uniformity & $\ns \leq O \left( \frac{\ab \log \ab}{\membits \dst^4}
    \right)$ & $\ns \geq \Omega \left( \frac{\ab \log \ab}{\membits \dst^4}
    \right)$ & $\ns \geq \Omega \left( \frac{\ab}{\membits \dst^2} \right)$ &
    $\membits \geq \Omega (\log \frac{\ab}{\dst}) $\\
    \hline
    Conditions & $\ab^{0.9} \gg \membits \gg \log (\ab) / \dst^2$ & $\membits =
    \tilde{\Omega} \left( \frac{\ab^{0.34}}{\dst^{8 / 3}} +
    \frac{\ab^{0.1}}{\dst^4} \right)$ & Unconditional & Unconditional \\    
    \hline
    & \multicolumn{4}{c|}{\textbf{Our Sample Complexity Bounds}} \\
    \hline
    \multirow{2}*{\textbf{Property}} & {\textbf{Upper Bound 1}} &
    {\textbf{Upper Bound 2}} & \multirow{2}*{\par\noindent\rule{1em}{1pt}} & 
    \multirow{2}*{\par\noindent\rule{1em}{1pt}} \\
    & (\cref{theo:theorem1}) & (\cref{theo:theorem2}) & & \\
    \hline
    Uniformity & $\ns \leq O \left( \frac{\ab \log \ab}{\membits \dst^4} \right)$ & $\frac{\ns}{\sqrt{\log \ns}} \leq O \left(
    \frac{\ab}{\sqrt{\membits} \dst^2} \right)$ & 
    \par\noindent\rule{1em}{1pt} & \par\noindent\rule{1em}{1pt} \\
    \hline
    Conditions & $\membits \leq \ab \log \ab$& & \par\noindent\rule{1em}{1pt} &
    \par\noindent\rule{1em}{1pt} \\
    \hline
  \end{tabular}
  \caption{Uniformity testing sample complexity with memory constraints. For our results (Upper bound 1 and 2) in the table, we implicitly assume that $\membits \geq \max (\log \ab, \log \ns, \log (1/\dst))$.  Our restriction $\membits \leq \ab\log\ab$ in the first column can be removed with relatively low effort; see~\cref{rk:removing:restriction}.\label{table:sample_complexity_uniformity}}
\end{table}

\begin{table}[ht]
  \centering
  \begin{tabular}{|c|c|c|}
    \hline
    \multirow{2}*{\textbf{property}} & \textbf{Prior Upper Bound} & \textbf{Our Upper Bound}\\
                                     & \cite{DiakonikolasGKR19} & (\cref{theo:theorem3})\\
    \hline
    Closeness & $O \Paren{ \frac{\ab \sqrt{\log (\ab)}}{\dst^2\sqrt{\membits}} }$ &
    $O \Paren{\frac{\ab\sqrt{\log \ns}}{\dst^2\sqrt{\membits}} + \frac{\ab^{2/3}}{\dst^{4/3}} }$\\
    \hline
    Conditions & $\tilde{\Theta}(\min(\ab, \frac{\ab^{2/3}}{\dst^{4/3}})) \gg \membits \gg \log \ab$ &\\
    \hline
  \end{tabular}
  \caption{Closeness testing sample complexity with memory constraints. The lower bound 1, 2 and 3 in \cref{table:sample_complexity_uniformity} still applies as this is a harder problem than uniformity testing. \label{table:sample_complexity_closeness}}
\end{table}

In a slightly orthogonal fashion, Berg, Ordentlich, and Shayevitz recently focused in~\cite{BergOS22} on the memory complexity of uniformity testing, regardless of the sample complexity (that is, even when the number of samples $\ns$ is allowed to grow unbounded). They provide (additive) bounds on the number of bits $\membits$ necessary and sufficient, as a function of $\ab,\dst$~--~while incomparable to our results, which focus on the \emph{tradeoff} between $\ns$ and $\membits$, theirs do imply a lower bound $\membits = \bigOmega{\log\ab + \log(1/\dst)}$ on the memory used by \emph{any} uniformity testing algorithm.

Finally, we contrast our results with those in the communication-constrained setting~\cite{FMO:18,ACT:19,ACLST:20}, where tight bounds on uniformity testing have been obtained under ``local'' communication constraints, namely where only $\numbits$ bits of communication can be sent about each of the $\ns$ samples. While the setting sounds similar, we note that the correspondence to memory constraint is very loose since the central server, in the communication-constrained setting, has no memory constraints and can store all $\ns\numbits$ bits of information received. That is, upper bounds in the communication-constrained setting would apply with $\membits = \ns\numbits$, while lower bounds only apply with $\membits = \numbits$ (the bottleneck in communication being $\numbits$ bits per sample). The tight sample complexity bound of $\bigTheta{\ab/(2^{\numbits/2}\dst^2)}$ for the communication-constrained setting~\cite{ACT:19,ACLST:20}, as a result, does not provide any meaningful bound in the streaming one.

Concurrent to our work, a recent paper by Roy and Vasudev~\cite{RoyV23} considers distribution testing of a range of properties in the streaming model. While relevant, we note that their results are orthogonal to ours, as they rely on the previous work of~\cite{DiakonikolasGKR19} on uniformity testing to obtain streaming algorithms for other properties than uniformity, using the (non-streaming) framework of~\cite{CDGR:17:journal,FLV:16} for ``shape-restricted properties.'' (They also consider streaming distribution testing in other access models than the standard \iid sampling one, specifically the conditional sampling model~\cite{CRS:15,CFGM:13}.) It would be interesting to see if our improvements upon~\cite{DiakonikolasGKR19} translate to better parameter regimes for the shape-restricted property testing results of~\cite{RoyV23}.

\subsection{Our results} The main contribution of our work is to provide conceptually simple algorithms, with elementary and concise proofs, which match the bounds of~\cite{DiakonikolasGKR19} while removing some (or, even, most) of the restrictions on the parameter regimes. Specifically, we obtain the following results:
\begin{theorem}
    \label{theo:theorem1}
There exists a (deterministic) one-pass streaming algorithm (\cref{algo:batch_streaming}) which, on input $\ab$ and $\dst\in(0,1]$, performs uniformity testing over $[\ab]$ using $\membits$ bits of memory and a stream of $\ns$ samples, as long as
\[
         \ns \geq C\cdot \frac{\ab\log \ab}{\dst^4\membits}
\]
and $\log\ab \leq \membits \leq \min(\ns\log\ab, \ab\log\ab)$, where $C>0$ is an absolute constant.
\end{theorem}
This first result is given by a deterministic algorithm. Our second result shows that, when allowing for randomization, we can obtain a different trade-off between $\ns$ and $\membits$, better (roughly) in the very low memory setting, when $\membits \ll (\log\ab)/\dst^4$.
\begin{theorem}
    \label{theo:theorem2}
There exists a (randomized) one-pass streaming algorithm (\cref{alg:hash_uniform_test}) which, on input $\ab$ and $\dst\in(0,1]$, performs uniformity testing over $[\ab]$ using $\membits$ bits of memory and a stream of $\ns$ samples, as long as
\[
         \ns \geq C'\cdot \frac{\ab\sqrt{\log \ns}}{\dst^2\sqrt{\membits}}
\]
and $\log\ab \leq \membits \leq \ab\log\ns$, where $C'>0$ is an absolute constant.
\end{theorem}
Finally, the ideas behind~\cref{theo:theorem2} straightforwardly extend to closeness testing, giving our third (and final) result:
\begin{theorem}
    \label{theo:theorem3}
There exists a (randomized) one-pass streaming algorithm which, on input $\ab$ and $\dst\in(0,1]$, performs closeness testing over $[\ab]$ using $\membits$ bits of memory and a stream of $\ns$ samples, as long as
\[
         \ns \geq C'\cdot \max\Paren{ \frac{\ab\sqrt{\log \ns}}{\dst^2\sqrt{\membits}}, \frac{\ab^{2/3}}{\dst^{4/3}} }
\]
and $\log\ab \leq \membits \leq \ab\log\ns$, where $C'>0$ is an absolute constant.
\end{theorem}

It is worth pointing out that plugging $\membits = \ns\log \ab$ in~\cref{theo:theorem1} retrieves the optimal sample complexity for uniformity testing in the unconstrained setting, $\ns = O(\sqrt{\ab}/\dst^2)$; while plugging $\membits = \ab\log \ns$ in~\cref{theo:theorem2} and~\cref{theo:theorem3} yields the optimal unconstrained sample complexity for uniformity and closeness testing, respectively.

\subsection{Outline of techniques}
Our starting point is the following obvious observation: encoding a set of $\ns$ samples over a domain of size $\ab$ can be done in two naive ways: (1)~first, the straightforward lossless encoding, which takes $O(\ns\log \ab)$ bits; and (2)~only keeping the counts (histogram), \ie the number of times each domain element is seen among the $\ns$ samples, which takes $O(\ab\log \ns)$ bits. (While the second option does lose some information, it is sufficient for any testing or learning question from i.i.d.\ samples, as the ordering of the $\ns$ samples does not matter.)

Now, given $\membits \ll \min(\ns\log\ab, \ab\log\ns)$ bits of available memory, we have two ``obvious'' options: either reduce the number of samples $\ns$, or reduce the domain size $\ab$, so that one of the two possible encodings fits into memory.

This, of course, seems \emph{a priori} hopeless, since we information-theoretically \emph{need} $\ns = \Omega(\sqrt{\ab}/\dst^2)$ samples for uniformity testing, and we do not get to choose the domain size. Yet, as we will see, by being careful (and a simple combination of existing ideas and tools from prior work), both strategies \emph{can} be implemented, and lead to painless algorithms matching the state-of-the-art.

\begin{itemize}
    \item Our first algorithm relies on a uniformity testing algorithm due to Diakonikolas, Gouleakis, Peebles, and Price~\cite{DGPP:18}, which happens to rely on a statistic $Z$ taking a very simple and convenient form when $\ns \ll \ab$~--~a form, in fact, which allows to compute and maintain $\nst$ independent copies of $Z$ (each on $\nss \ll \ns$ samples) using only $O(\log(\nst\ab))$ bits in total. The key is then to compute $\nst$ such statistics on $\nss$ samples each and average them at the end, leading to a sample complexity $\ns = \nss\cdot\nst$ using $\membits = \nss\log\ab +\log(\nst\ab)$ bits of memory. Balancing the two, along with straightforward analysis of the expectation and variance of the average of these $\nst$ copies, yields~\cref{theo:theorem1}.
    \item Our second algorithm relies on the primitive of \emph{domain compression} introduced by Acharya, Canonne, and Tyagi~\cite{ACT:19}, a variant of hashing tailored to distribution testing and learning which (roughly) allows one to trade domain size $\ab$ for error parameter $\dst$. That is, one can transform an instance of testing over domain size $\ab$ and distance parameter $\dst$ to a new instance over domain size $\ab'$ and (smaller) distance parameter $\dst' \asymp \dst \sqrt{\ab'/\ab}$. Setting $\ab' = \membits/\log\ns$, now the memory can fit all sample counts! And since we are now performing uniformity testing (with full information) with domain $[\ab']$ and  parameter $\dst'$, the resulting testing algorithm works as long as the number of samples $\ns$ satisfies
    \[
            \ns \gtrsim \frac{\sqrt{\ab'}}{\dst'^2}
    \]
    which, recalling the setting of $\ab',\dst'$, simplifies to the desired result,
    \[
            \ns \gtrsim \frac{\ab\sqrt{\log \ns}}{\dst^2\sqrt{\membits}}\,,
    \]
    and (give or take a few additional details) establishes~\cref{theo:theorem2}.
\end{itemize}
One additional feature of the second strategy is that it does not, in fact, rely on anything specific to uniformity testing at all, besides invoking an out-of-the-box ``standard'' uniformity testing algorithm at the very end on the sample counts of the ``reduced'' instance. But one could apply the exact same idea~--~domain compression to fit the sample counts in memory before using an out-of-the-box algorithm on them at the end of the stream~--~to other distribution testing problems: the end result will then only depend on the sample complexity of this testing problem, when applied to the parameters $\ab',\dst'$ obtained after domain compression. This is exactly what we do in~\cref{ssec:closeness} to  obtain our closeness testing streaming result,~\cref{theo:theorem3}.

\section{Testing via Repetition}
Our first algorithm, whose analysis will establish~\cref{theo:theorem1}, will rely upon the uniformity testing algorithm of Diakonikolas, Gouleakis, Peebles, and Price~\cite{DGPP:18}, which works as follows: given $\ns$ \iid samples from the unknown distribution $\p$ over $[\ab]$, let $\freq_1,\dots, \freq_{\ab}$ denote the corresponding counts (so that $\sum_{i=1}^{\ab} \freq_i = \ns$), and consider the quantity
\begin{equation}
    \label{def:Z:dgpp}
    Z \eqdef \frac{1}{2}\sum_{i=1}^{\ab} \abs{\frac{\freq_i}{\ns} - \frac{1}{\ab}}
\end{equation}
which corresponds to the (total variation) distance to uniform of the empirical distribution obtained. The main contribution of~\cite{DGPP:18} is to show that comparing $Z$ to a suitable threshold $\tau=\tau(\ns,\dst,\ab)$ does, in fact, lead to a sample-optimal uniformity testing algorithm.

While it is not clear \emph{a priori} how this would help in the memory-limited setting, one nice feature of this quantity $Z$ is that the slightly unwieldy expression in~\eqref{def:Z:dgpp} simplifies to a much nicer form when $\ns\leq \ab$: namely, since then $\abs{\freq_i - \frac{\ns}{\ab}}$ is either $\frac{\ns}{\ab}$ if $\freq_i=0$ or $\freq_i - \frac{\ns}{\ab}$ if $\freq_i\neq 0$ ($\freq_i$ being an integer), one can easily check that\todonote{One can also encode the original $Z$ with $\log (k n)$-bits, as $n k Z$ is at most $n k$. Or it is $\log (s k)$ in batches.}
\begin{equation}
    \label{def:Z:dgpp:nlessk}
    Z  = \frac{1}{\ab}\sum_{i=1}^{\ab} \indicSet{\freq_i = 0}
\end{equation}
\ie $Z$ now is just the (normalized) number of \emph{unseen} elements of the domain~--~which, once computed, only takes $\log_2(\ab+1)$ bits to store! What's even better, storing the running average of $\nst$ independent copies $Z_1,\dots, Z_{\nst}$ of $Z$ only takes $\log_2 \nst + \log_2(\ab+1)$ bits, as $\nst\ab\cdot(Z_1 + \dots + Z_{\nst})$ is simply an integer in $\{0,1,\dots,\nst\ab\}$.

However, computing even \emph{one} copy of $Z$ from $\ns$ samples takes (at least when done naively) memory roughly $\ns\log\ab$ (or alternatively $\ab$, which is either worse or not much better in our regime $\ns \leq \ab$), keeping in memory all samples. It seems that we are back to square one!

Fortunately, there is a simple fix to this: divide the stream of $\ns$ independent samples into $\nst$ batches of $\nss$ samples, and compute one independent $Z_t$ per batch $t$, on only the $\nss$ samples from this batch that we then only have to keep in memory during the current batch, and can discard afterwards. By choosing $\nss$ so that $\membits \geq \nss \log \ab$, we can afford to do so; and keeping track (once $Z_t$ is computed) of the running sum $Z_1+\dots+Z_t$, we will be able to average the resulting $\nst$ values $Z_1,\dots, Z_{\nst}$ using only an additional $\log_2 \nst + \log_2(\ab+1)$ bits. That is, we need to choose $\nss,\nst$ so that
\begin{equation}
    \label{eq:setting:m:alg1}
    \membits \geq \nss \log_2 \ab + \log_2 \nst + \log_2(\ab+1) 
\end{equation}
after which, at the end of the stream, it will be enough to threshold the average $\frac{1}{\nst}(Z_1+\dots+ Z_{\nst})$ at the value $\tau = \tau(\nss,\dst,\ab)$.

Note that having $\membits \gg \log \ns$ (by~\eqref{eq:relevant:regime:m}) and $\nst\leq \ns$ imply that $\log \nst \ll \membits$, and as a result choosing $\nss,\nst$ according to~\cref{eq:setting:m:alg1} will lead to $\membits = \Theta(\nss\log\ab)$. Importantly, our condition $\membits \ll \ab\log\ab$ then implies $\nss \leq \ab$, which we need for~\eqref{def:Z:dgpp:nlessk} to hold.

To conclude, it ``only'' remains to argue correctness: that is, to establish (1)~that each $Z_t$ has an expectation noticeably different under the uniform distribution and under a distribution that is $\dst$-far from uniform, and (2)~the number of batches $\nst$ needed for the averaging to concentrate well enough around that expectation, so that the thresholding yields the right answer with probability at least $2/3$. Thankfully, this has already been taken care of! Using the analysis of~\cite{DGPP:18} (as slightly simplified/modified in~\cite[Section~2.1.5]{CanonneTopicsDT2022} for the regime $\ns \leq \ab$, and to get a variance bound), we have that, computing $Z$ from $\nss$ samples, for every $\p$ that is $\dst$-far from the uniform distribution $\uniform_{\ab}$, the gap in expectation is
\begin{equation}
    \bE{\p}{Z}-\bE{\uniform_{\ab}}{Z} \geq \frac{\nss^2\dst^2}{4e\ab^2} \coloneqq \Delta
\end{equation}
while the variance of $Z$ is at most
\begin{equation}
    \label{eq:variance:bound:empirical}
    \var_{\uniform_{\ab}}[Z], \var_{\p}[Z] \leq \frac{2\nss^2}{\ab^3}
\end{equation}
(see~\cite[Eqs~(2.28) and (2.35)]{CanonneTopicsDT2022}). Averaging over our $\nst$ independent copies, the gap in expectation $\Delta$ remains, but the variance drops by a factor $\nst$: letting $\bar{Z} \eqdef \frac{1}{\nst}(Z_1+\dots+Z_{\nst})$,
\begin{equation}
    \label{eq:algo1:var}
   \bE{\p}{\bar{Z}} -  \bE{\uniform_{\ab}}{\bar{Z}}  \geq \Delta, \qquad \var[\bar{Z}] \leq \frac{2\nss^2}{\nst\cdot \ab^3}
\end{equation}
For the value of the threshold
\begin{equation}
    \label{eq:algo1:tau}
    \tau \eqdef \bE{\uniform_{\ab}}{\bar{Z}} + \frac{\Delta}{2} = \Paren{1-\frac{1}{\ab}}^{\nss} + \frac{\nss^2\dst^2}{8e\ab^2}
\end{equation}
we get, by Chebyshev, that the probability that the algorithm errs is at most, both under the uniform and far-from-uniform cases,
\begin{equation}
    \bPr{\abs{\bar{Z} - \bEE{\bar{Z}} } \geq \frac{\Delta}{2}  } \leq \frac{4\var[\bar{Z}]}{\Delta^2} \leq \frac{512e^2}{\nst}\cdot \frac{\ab}{\nss^2\dst^4}
\end{equation}
using~\eqref{eq:algo1:var} and~\eqref{eq:algo1:tau}; this is at most $1/3$ for $\nst \geq 1536e^2\cdot \frac{\ab}{\nss^2\dst^4}$. Put differently, the algorithm works as long as $\nss^2\nst \geq 1536e^2\cdot \ab/\dst^4$; recalling that $\ns = \nss\nst$ and (from~\eqref{eq:setting:m:alg1}) that $\membits = \Theta(\nss\log\ab)$, we get that it is enough to have $\frac{\ns\membits}{\log\ab} \geq C\cdot \ab/\dst^4$ for some absolute constant $C>0$, proving~\cref{theo:theorem1}.

\begin{remark}[We lied] The above argument glosses over a technical detail, which, while innocuous, needs to be addressed: namely, that the variance bound given in~\eqref{eq:variance:bound:empirical} only holds for \emph{some} of the $\dst$-far distributions $\p$, those with small $\lp[\infty]$ norm. These are, in a quantitative sense, the worst-case instances for the algorithm, as shown in~\cite{DGPP:18} \emph{via} stochastic dominance,\footnote{We refer the reader to either~\cite{DGPP:18} or~\cite[Section~2.1.5]{CanonneTopicsDT2022} for the formal definition, and a discussion. Note that the variance bound~\eqref{eq:variance:bound:empirical} fails to hold for some ``easy-looking'' distributions such as, \eg{} a distribution uniform on a subset of $\nss \ll \ab$ elements, for which the variance becomes $\bigTheta{\nss/\ab^2}$. But while this distribution leads to a much worse variance, it also comes with a much larger expectation gap, so overall is, indeed, ``easier.''} and thus it suffices to consider these particular distributions. This also applies to our case, as we consider an average of these statistics $Z_1,\dots,Z_{\nst}$, and thus the same stochastic dominance argument goes through. 
\end{remark}

\begin{remark}[What about $\nss \gg \ab$?]
    \label{rk:removing:restriction}
    To lift the restriction on $\nss \leq \ab$ (or equivalently $\membits < \ab \log \ab$), one can consider the empirical distance tester in \eqref{def:Z:dgpp} for the regime $\nss > \ab$ and note that each batch of $Z$ can be represented in memory by storing $\sum_{i=1}^{\ab} \abs{\freq_i \ab - \nss}$ and $\nss \ab$, which uses at most $\log (2 \nss \ab \nst) = \log (2 \ns \ab)$ memory over $\nst$ rounds. Thus the memory used in total remains $\membits = \Theta (\nss \log \ab) = \Theta (\nss \log \ab + \log (\nss \ab))$. By analyzing the variance of \eqref{def:Z:dgpp}, which in the regime $\nss > \ab$ is $O(1/\nss)$ (this follows from~\cite[Section~2.3.1]{DGPP:19}) along with the expectation gap (which is $\Omega(\dst^2\sqrt{\nss/\ab})$ for $\ab < \nss\leq \ab/\dst^2$ and $\Omega(\dst)$ for $\nss > \ab/\dst^2$~\cite[Lemma~4]{DGPP:19}), we obtain an unrestricted version of the batch streaming tester with the same memory-sample complexity trade-off.
\end{remark}

\begin{remark}[What about the collision-based tester?]
    \label{rk:using:collisionbased}
Our choice of using the empirical total variation distance tester of~\cite{DGPP:18} (given in~\eqref{def:Z:dgpp}) may seem a little arbitrary: we essentially chose it for the simple form it takes in the regime $\nss \ll \ab$, as well as its additional generalization properties coming from its low sensitivity with respect to the samples, which we believe could come handy for future work (\eg{} for robustness, privacy, and high-probability testing). However, within our ``testing via repetition'' streaming framework, one could use the collision-based tester instead~\cite{GRexp:00,DGPP:19}, which similarly only requires to keep a counter in each batch (for the number of collisions). We leave as an exercise to the interested reader to show that this would, indeed, result in the bound. Importantly, this would not make the argument simpler, due to the subtlety in the variance analysis of that tester necessary to get the right dependence on $\dst$, for which one would have to rely on the analysis of~\cite{DGPP:19} (see also~\cite[Section~2.1.2]{CanonneTopicsDT2022}).
\end{remark}

\begin{algorithm}[htb]
\caption{Uniformity testing in batches\label{algo:batch_streaming}}
\begin{algorithmic}[1]
    \Statex \textbf{Input}: stream of $\ns$ samples from distribution $\p$, accuracy $\dst$, domain size $\ab$, memory bound $\membits$
    \State $\nss \gets \Theta (\membits/\log \ab)$, $\nst \gets \frac{\ns}{\nss}$ as in~\eqref{eq:setting:m:alg1}; %
    \State $\tau \gets \Paren{1-\frac{1}{\ab}}^{\nss} + \frac{\nss^2\dst^2}{8e\ab^2}$;
    \State $Z \gets 0$;
    \For{$i=1$ to $\nst$}
      \State $\mathcal{S} \gets$ $\nss$ samples; \Comment{$O(\nss\log\ab)$ bits}
      \State Obtain $\freq_1, \ldots, \freq_{\ab}$ from $\mathcal{S}$;
      \State $Z_i \gets \frac{1}{\ab}\sum_{j=1}^{\ab} \indicSet{\freq_j = 0}$;
    \EndFor
    \State $Z \gets \frac{1}{\nst} \sum_{j=1}^{\nst} Z_j$;
    \If {$Z > \tau$} {
      \reject
    } \Else {
      \accept
    }
    \EndIf
\end{algorithmic}
\end{algorithm}

\section{Testing via Domain Compression}

We will rely on the following theorem from previous work, which provides the ``domain compression'' primitive:\footnote{We here use the domain compression lemma with respect to total variation ($\lp[1]$) distance; for the weaker, but sufficient $\lp[2]$ version, one could instead invoke~\cite[Theorem~VI.2]{ACT:19}.}
\begin{theorem}[{\cite[Theorem~5]{ACHST:20}; see also~\cite[Theorem~2.12]{CanonneTopicsDT2022}}]
    \label{lemma:dct}
    There exist absolute constants $c_1, c_2 > 0$ such that the following holds. For any $2 \leq \ab' \leq \ab$ and any two distributions $\p,\q$ over $[\ab]$,
    \[
        \probaDistrOf{\Pi}{\totalvardist{\p_\Pi}{\q_\Pi} \geq c_1\cdot \sqrt{\frac{\ab'}{\ab}}\totalvardist{\p}{\q}} \geq c_2
    \]
where $\Pi$ is a random partition of $[\ab]$ in $\ab'$ subsets,
and $\p_\Pi$ denotes the probability distribution over $[\ab']$ induced by $\p$ and $\Pi$ via $\p_\Pi(i)=\p(\Pi_i)$. Moreover, $\Pi$ can be sampled and encoded using $O(\log\ab)$ bits.
\end{theorem}
One can equivalently see the random partition $\Pi$ from the above theorem as a hash function $\pi\colon[\ab]\to[\ab']$ represented by $O(\log\ab)$ random bits. With this succinct representation, given a sample $X$ from $\p$ (over $[\ab]$) one can then compute the induced sample $X'$ from $\p_\Pi$: $X'=\pi(X) \in [\ab']$.

Armed with the above, set $\ab'\geq 2$ to be the largest integer such that
\begin{equation}
    \label{eq:hashing:m}
    \membits = \underbrace{\Theta(\log\ab)}_{\rm(1)} + \underbrace{\Theta(\ab'\log \ns)}_{\rm(2)}
\end{equation}
bits are enough to encode (1)~the random bits succinctly encoding the partition $\Pi$, and (2)~the counts from the $\ns$ induced samples from $\p_\Pi$. In view of~\eqref{eq:relevant:regime:m}, this leads to $\membits = \Theta(\ab'\log \ns)$, and in particular $\membits \leq c_3 \cdot \ab'\log \ns$ (where $c_3>0$ is an absolute constant).\footnote{In particular, in view of the restriction $\ab' \leq \ab$ (one cannot compress the domain to a \emph{larger} domain), we have the restriction $\membits \ll \ab\log\ns$, consistent with~\eqref{eq:relevant:regime:m}.} 
Further, set
\begin{equation}
    \label{eq:hashing:eps}
    \dst' = c_1\cdot \sqrt{\frac{\ab'}{\ab}} \cdot \dst\,,
\end{equation}
where $c_1>0$ is the constant from~\cref{lemma:dct}. This allows us to keep in memory the counts (histogram) corresponding to the $\ns$ samples from our induced distribution $\p_\Pi$ over $[\ab']$, and therefore by~\cref{lemma:dct} all we need to do now is gather enough samples to solve the uniformity testing question over domain $[\ab']$ with distance parameter $\dst'$, using any algorithm which only requires the counts: \eg the sample-optimal $\chi^2$-based tester of~\cite{AcharyaDK15} (see also~\cite[Theorem~2.5]{CanonneTopicsDT2022}). 

Before analyzing the number of samples $\ns$ that suffice for this, we need to address one possible wrinkle: namely, that the guarantee of~\cref{lemma:dct} only holds with (small) constant probability $c_2$ over the choice of $\Pi$.\footnote{Note that if $\p=\q$, then of course $\p_\Pi=\q_\Pi$ with probability one, so the only amplification needed is to make sure that we do get $\totalvardist{\p_\Pi}{\q_\Pi} > \dst'$ some time, when $\totalvardist{\p}{\q} > \dst$.} This is not a serious problem however, as one can amplify this probability $c_2$ to any constant arbitrarily close to one via a standard amplification argument, at the cost of a constant number of sequential independent repetitions (constant factor loss in the resulting sample complexity) and maintaining a counter for these repetitions (at the cost of a constant overhead in the memory complexity); for the sake of completeness, we recall this standard argument in~\cref{app:standard:amplification}. Thus, we ignore this constant-factor and additive-constant overheads in the remainder of the proof.

To have a successful algorithm, recalling the standard sample complexity of uniformity testing, it suffices for $\ns$ to satisfy
\begin{equation}
    \label{eq:bound:nk':eps'}
    \ns \geq c_4\cdot \frac{\sqrt{\ab'}}{\dst'^2}
\end{equation}
where $c_4>0$ is (yet another) absolute constant. From our settings of $\ab'$ and $\dst'$ from~\eqref{eq:hashing:m} and~\eqref{eq:hashing:eps}, this gives that having 
\begin{equation}
    \ns \geq c_1^{-1} c_3^{1/2} c_4\cdot \frac{\ab\sqrt{\log \ns}}{\dst^2\sqrt{\membits}}
\end{equation}
is sufficient. This proves~\cref{theo:theorem2}.

\begin{algorithm}[htb]
\caption{Uniformity testing via domain compression\label{alg:hash_uniform_test}}
\begin{algorithmic}[1]
    \Statex \textbf{Input}: stream of $\ns$ samples from distribution $\p$, accuracy $\dst$, domain size $\ab$, memory bound $\membits$
    \State Set $\ab' \leftarrow \Theta \Paren {\frac{\membits}{\log \ns}}, \dst' \leftarrow c_1\cdot \sqrt{\frac{\ab'}{\ab}} \cdot \dst$ and $\errprob \gets 1/3$;
    \State Get hash function $\pi$ as in \cref{lemma:dct};
    \For{$i=1$ to $\ns$}
        \State Hash $i$-th sample $x_i$ to $x'_i\gets \pi(x_i)$, and keep $x'_i$ in memory;
    \EndFor
    \State Run any sample-optimal uniformity testing algorithm on $x'_i,\dots,x'_\ns$, with parameters $\dst',\ab',\errprob$.
    \If {That algorithm rejects} {
      \reject
    } \Else {
      \accept
    }
    \EndIf
\end{algorithmic}
\end{algorithm}

\subsection{Closeness testing}
    \label{ssec:closeness}
In the previous section, we provided a one-pass uniformity testing algorithm based on domain compression. The reader may have noticed that most of this algorithm does not, in fact, rely on anything else that the fact domain compression preserves distances between distributions, and has nothing specific to uniformity testing except for the very last step (where a ``standard'' uniformity testing algorithm algorithm is invoked on the $\ns$ induced samples on the compressed domain). It is straightforward to extend~\cref{algo:batch_streaming} to other distribution testing problems, and in particular the (related) question of \emph{closeness} testing, where instead of a stream of $\ns$ \iid samples from one unknown distribution $\p$, one gets $\ns$ samples from two unknown distributions, $\p$ and $\q$, and the goal is to test whether $\p=\q$ or $\totalvardist{\p}{\q}>\dst$. The only modification to~\cref{algo:batch_streaming} will be to invoke a sample-optimal ``standard'' \emph{closeness} testing algorithm at the end, still with parameters $\ab',\dst'$. Since the optimal sample complexity of closeness testing is $\Theta(\max(\sqrt{\ab}/\dst^2, \ab^{2/3}/\dst^{4/3}))$~\cite{ChanDVV14,DK:16,DiakonikolasGKPP21,CanonneS22}, the analogue of~\eqref{eq:bound:nk':eps'} becomes
\begin{equation}
    \ns \geq c_4\Paren{\frac{\sqrt{\ab'}}{\dst'^2} + \frac{\ab'^{2/3}}{\dst'^{4/3}} }
\end{equation}
leading to the sufficient condition
\begin{equation}
    \ns \geq c' \Paren{\frac{\ab\sqrt{\log \ns}}{\dst^2\sqrt{\membits}} + \frac{\ab^{2/3}}{\dst^{4/3}} }
\end{equation}
for some absolute constant $c'>0$: this proves~\cref{theo:theorem3}.

\section{Discussion and future work}
We note that our results leave open a few interesting directions. The first, and most obvious, is whether one can improve the \emph{lower bounds} of~\cite{DiakonikolasGKR19} to prove optimality of the tradeoffs obtained, in all parameter regimes. The second is to extend the general outline used in~\cref{theo:theorem2,theo:theorem3} to other testing problems: that is, which other distribution testing questions are amenable to efficient streaming algorithms \emph{via} domain compression? 

Finally, recall that our first algorithm is deterministic, while the second relies on domain compression (hashing), and is thus randomized. It would be interesting to study whether there exists, in some parameter regime, a separation between the power of deterministic and randomized algorithms for uniformity testing, as is the case under communication constraints~\cite{AcharyaCT:IT1,ACT:19}.

\todonoteinline{Combined use of the two algorithms. For uniformity testing, one can use a combination of the two algorithms presented in the paper}

\paragraph{Acknowledgments.} We thank the anonymous reviewers of the SIAM Symposium on Simplicity in Algorithms (SOSA24) for their helpful comments and suggestions, which among others led to~\cref{rk:removing:restriction,rk:using:collisionbased}.

\printbibliography

\appendix
\section{About the standard amplification trick}
\label{app:standard:amplification}
We can leverage the fact that in uniformity testing, uniform distribution being mapped to a smaller domain will remain uniform (on a smaller domain) with probability one.\footnote{We can use the same idea in closeness testing: mapping two distributions that are the same to any smaller domain does not affect the $\operatorname{TV}$ distance in the completeness case.} Because of this, any hashing is good in the case that $\p$ is uniform (in the completeness case). Suppose we have a uniformity testing algorithm that is correct except with some (sufficiently small, to be determined) probability $\errprob$. If $\p=\uniform_{\ab}$, then after hashing the induced distribution on $[\ab']$ will be accepted with probability at least $1-\errprob$.

Meanwhile, when $\p$ is $\dst$-far from uniform (in the soundness case), the mapping is good with probability at least $c_2>0$ and the tester will thus reject with probability at least $(1-\errprob) c_2$; \ie the tester will accept with probability at most $1 - (1 - \errprob) c_2$. To be able to amplify by repetition, we need a gap between the two acceptance probabilities of the two cases: 
\[
    1-\errprob \gg 1 - (1-\errprob) c_2
\]
which is satisfied for any choice of $\errprob < \frac{c_2}{1+c_2}$ (note that this is a constant). By taking some constant (determined by this choice of $\errprob$, that is, by $c_2$) repetition and comparing the average acceptance rate against the threshold $1 - \frac{\errprob + (1-\errprob) c_2}{2}$, one can separate the two cases (with Chernoff bound) with probability at least $2/3$.
\end{document}